\documentclass[usenatbib,useAMS,letters]{mnras}

\usepackage{natbib}
\usepackage{amsmath}
\usepackage{graphicx}
\usepackage{color}
\usepackage{mathrsfs}
\usepackage{epsfig}
\usepackage{comment}
\usepackage{ulem}

\usepackage{graphicx}
\usepackage[hang,small,bf]{caption}
\usepackage[subrefformat=parens]{subcaption}
\usepackage{amssymb}
\usepackage[T1]{fontenc}

\newcommand{\msun}{{M}_{\odot}}
\newcommand{\Mc}{M_{\rm chirp}}
\newcommand{\Mt}{M_{\rm total}}

\newcommand{\beq}{\begin{equation}}
\newcommand{\eeq}{\end{equation}}

\defcitealias{K14}{K14}
\defcitealias{K16}{K16}
\defcitealias{Visbal_2015}{VHB15}

\usepackage{url} 

\title[Pop III BBHs are consistent with  O3a]
{Gravitational waves from Population III binary black holes are consistent with LIGO/Virgo O3a data for the chirp mass larger than $\sim 20\msun$ }

\author[T. Kinugawa et al.]
{Tomoya Kinugawa$^{(1)}$\thanks{E-mail: kinugawa@icrr.u-tokyo.ac.jp}, Takashi Nakamura$^{(2)}$, and Hiroyuki Nakano$^{(3)}$\\
\\
$^{1}$Institute for Cosmic Ray Research, The University of
  Tokyo, Kashiwa, Chiba 277-8582, Japan\\
$^{2}$Department of Physics, Graduate School of Science, Kyoto University,
Kyoto 606-8502, Japan\\
$^{3}$Faculty of Law, Ryukoku University, Kyoto 612-8577, Japan}

\begin{document}

\date{\today}
\maketitle


\begin{abstract}
The probability number distribution function of binary black hole mergers observed by LIGO/Virgo O3a has double peaks as a function of chirp mass $\Mc$, total mass $\Mt$, primary black hole mass $M_1$ and secondary one $M_2$, respectively. The larger chirp mass peak is at $\Mc \cong 30 \msun$. The distribution of $M_2$ vs. $M_1$ follows the relation of $M_2\cong 0.7M_1$. For initial mass functions of Population III stars in the form of $f(M) \propto M^{-\alpha}$, population synthesis numerical simulations with $0\leq \alpha \leq 1.5$ are consistent with O3a data for $\Mc \gtrsim 20\msun$. The distribution of $M_2$ vs. $M_1$ for simulation data also agrees with $M_2\cong 0.7M_1$ relation of O3a data. 
\end{abstract}


\section{Introduction}

\begin{figure}
    \centering
    \includegraphics[width=0.47\textwidth]{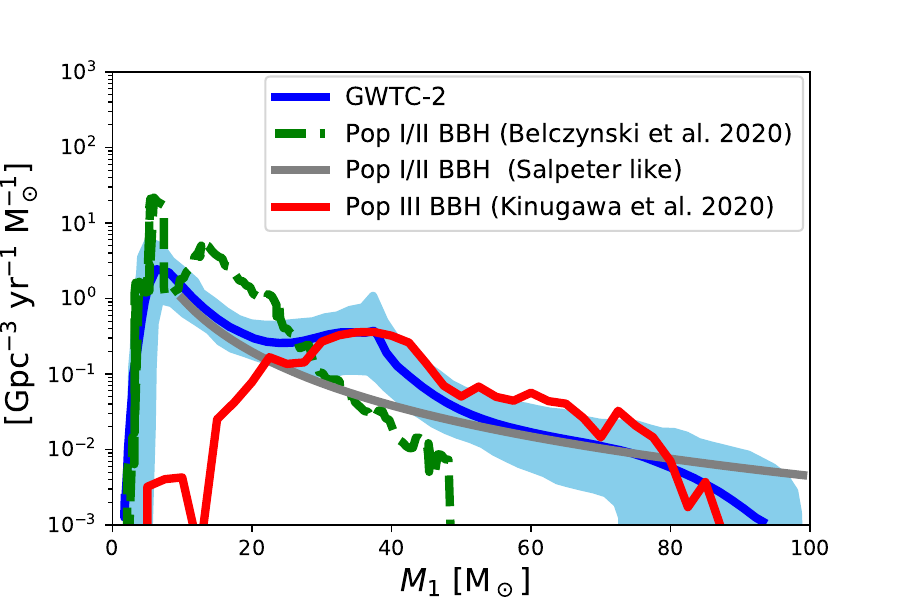}
    \caption{BBH merger rate density as a function of primary mass ($M_1$). The blue line and the light blue region are the population distribution and 90\% credible interval of the power law + peak model by GWTC-2, respectively~\citep{2020arXiv201014533T}. The green dashed line is the Pop I/II BBH merger rate density of the M30.B model in~\citet{Belczynski2020}.  {The gray line is the Salpeter like power law model ($M_1^{-2.35}$) for Pop I/II BBHs which is normalized by GWTC-2 rate at $M_1=10\msun$.} The red  line shows the Pop III BBH merger rate density of the M100 model in~\citet{Kinugawa2020} which is the same as the flat model in this paper.}
    \label{M1dist}
\end{figure}

The second LIGO--Virgo Gravitational-Wave Transient Catalog (GWTC-2) was announced on October 28, 2020~\citep{2020arXiv201014527A}.
In a companion paper~\citep{2020arXiv201014533T}, 
population properties of compact object binaries observed during the first half of the third observing run (O3a)
have been discussed, especially by focusing on the primary mass and spin distributions for BBHs (Binary Black Holes).
Specifically, they analysed the merger rate density distribution as a function of primary mass $M_1$, and showed that the power law + peak model is the most likely one.
Figure~\ref{M1dist} plots the power law + peak model with BH mass distribution of Population (Pop) I/II BH \citep{Belczynski2020} and Pop III BH \citep{Kinugawa2020}.
The power law + peak model looks like consistent with the power law mass distribution of Pop I/II BHs \citep{Belczynski2020} and the mass distribution of Pop III BHs with the peak at $\sim 30$--$40\msun$ \citep{Kinugawa2020}.
Since Figure~\ref{M1dist} is shown as a function of the primary mass $M_1$, it is also important to study the merger rate as functions of secondary mass $M_2$, total mass $M_{\rm total}$ and chirp mass $\Mc$ in order to check whether the Pop III BH model is consistent with  the peak of massive BBHs or not, varying IMF (Initial Mass Function) and SFR (Star Formation Rate).

In this Letter, firstly, we focus on the chirp mass distribution
of BBHs
because the chirp mass is the most sensitive parameter
in GW observation of compact object binaries where
the chirp mass ($\Mc$) is defined by
$\Mc = (M_1 M_2)^{3/5}/(M_1+M_2)^{1/5}$.
 Secondly, we omit binaries 
which include a compact object with mass $< 3\msun$, i.e., 
GW190425 ($M_1=2\msun$, $M_2=1.4\msun$),
GW190426\_152155 ($M_1=5.7\msun$, $M_2=1.5\msun$)
and GW190814 ($M_1=23.2\msun$, $M_2=2.59\msun$)
in the 39 GW events in GWTC-2 O3a
so that we consider  36 GW events only. Thirdly, we do not treat spins of BBHs
although there are interesting results 
 {related to the BH spins~\citep{2021MNRAS.tmp..279F,2020arXiv201109570C,2020arXiv201111948G,2021arXiv210201689T}.}
This is because there are still large uncertainties
in the estimation of BH spins.

 {After the announcement of GWTC-2, various interesting papers have been presented;
\cite{2020PhRvD.102l3016A} found a drop in the BBH merger rate for $M_1 \lesssim 13$ and $\gtrsim 30\msun$ in a population synthesis code of BBH formation in globular clusters (GCs) with a wide set of initial conditions (note that hierarchical mergers to create heavier primary mass BHs are only $\lesssim 10\%$ of the total number of BBH mergers expected from GCs~\citep{Rodriguez:2017pec,Rodriguez:2019huv}),
\cite{2020arXiv201104502T} reanalyzed GWTC-2 for the chirp mass, mass ratio, and spin distributions with minimal assumptions~\citep{2020arXiv200615047T} and found peaks in the chirp mass distribution at $8$, $14$, $26$, and $45\msun$,
\cite{2020arXiv201105332K} presented that GWTC-2 is best modelled with hierarchical formation channels by using a phenomenological population model~\citep{2020ApJ...900..177K} based on simulations of metal-poor GCs~\citep{Rodriguez:2019huv} (note that GWTC-1~\citep{2019PhRvX...9c1040A} was consistent with having no hierarchical merger in this model),
\cite{2021ApJ...907L..48V} gave a search for hierarchical triple mergers including spin effects, and analyzed the GW events by assuming upper bounds on the mass distribution of first generation BHs,
and \cite{2021arXiv210107699F} paid attention to the absence of BBH events with $M_1 > 45\msun$ at low redshifts, and discussed the evolution of the BBH mass distribution
that will be distinguishable in future GW observations.
\cite{2020arXiv201107000B} obtained 
BBH merger rate densities and differential BBH merger rate densities which are consistent with the LIGO--Virgo result,
for dynamical BBH formation in young massive star clusters
and open star clusters
based on N-body evolutionary models of star clusters including hierarchical mergers~\citep{2020MNRAS.500.3002B}.
Although \cite{2021arXiv210107793R} showed that the redshift-dependent merger rate of GWTC-2 can be explained by a purely dynamical origin in GCs, they cautioned that various formation scenarios could contribute the rate. In practice, \cite{2020arXiv201103564W,2020arXiv201110057Z,2021arXiv210212495B} have considered mixture models by introducing hyper-parameters to describe the fraction of each formation channel.
We also see scenarios with primordial BHs (PBHs) formed in the early Universe; \cite{2020arXiv201101865W} introduced a PBH scenario with accretion to explain the presence of several spinning BBHs in GWTC-2, \cite{2021arXiv210111098D} discussed a possible mass distribution of primordial black holes by assuming that all LIGO/Virgo BBHs have a primordial origin, and \cite{2020arXiv201202786H,2021arXiv210203809D} considered combination of populations of astrophysical BHs and PBHs (see also \cite{Hall:2020daa} for GWTC-1).}

{Pop III BHs are not included in any analysis of BBH formation models mentioned above, although the Pop III binaries are considered as a candidate of the massive BBH origin \cite[e.g.][]{Kinugawa2014,Kinugawa2020,Kinugawa2021,Tanikawa2020,Farrell2020}.}
We believe that we can add one more interesting paper
on the O3a events.

\section{Analysis}

\subsection{Data from LIGO/Virgo GWTC-2 O3a}

Using the median values and 90\% credible intervals
of parameters estimated by~\cite{2020arXiv201014527A},
we prepare a simple probability number distribution function
of various mass `$m$'
to introduce the parameter estimation errors as
\begin{align}
g(m) = &
(2\pi\sigma_-^2)^{-1/2} \exp\left\{-\frac{(m-m_0)^2}{2\sigma_-^2}\right\} \Theta(m_0-m)
\cr &
+
(2\pi\sigma_+^2)^{-1/2}  \exp\left\{-\frac{(m-m_0)^2}{2\sigma_+^2}\right\} \Theta(m-m_0) \,,
\label{eq:dist}
\end{align}
where each side of $m=m_0$
has 50\% probability, i.e., $m_0$ is the median, and $\sigma_{-}$ and $\sigma_+$
are determined from the 90\% credible interval.

\begin{figure}
    \centering
    \includegraphics[width=0.47\textwidth]{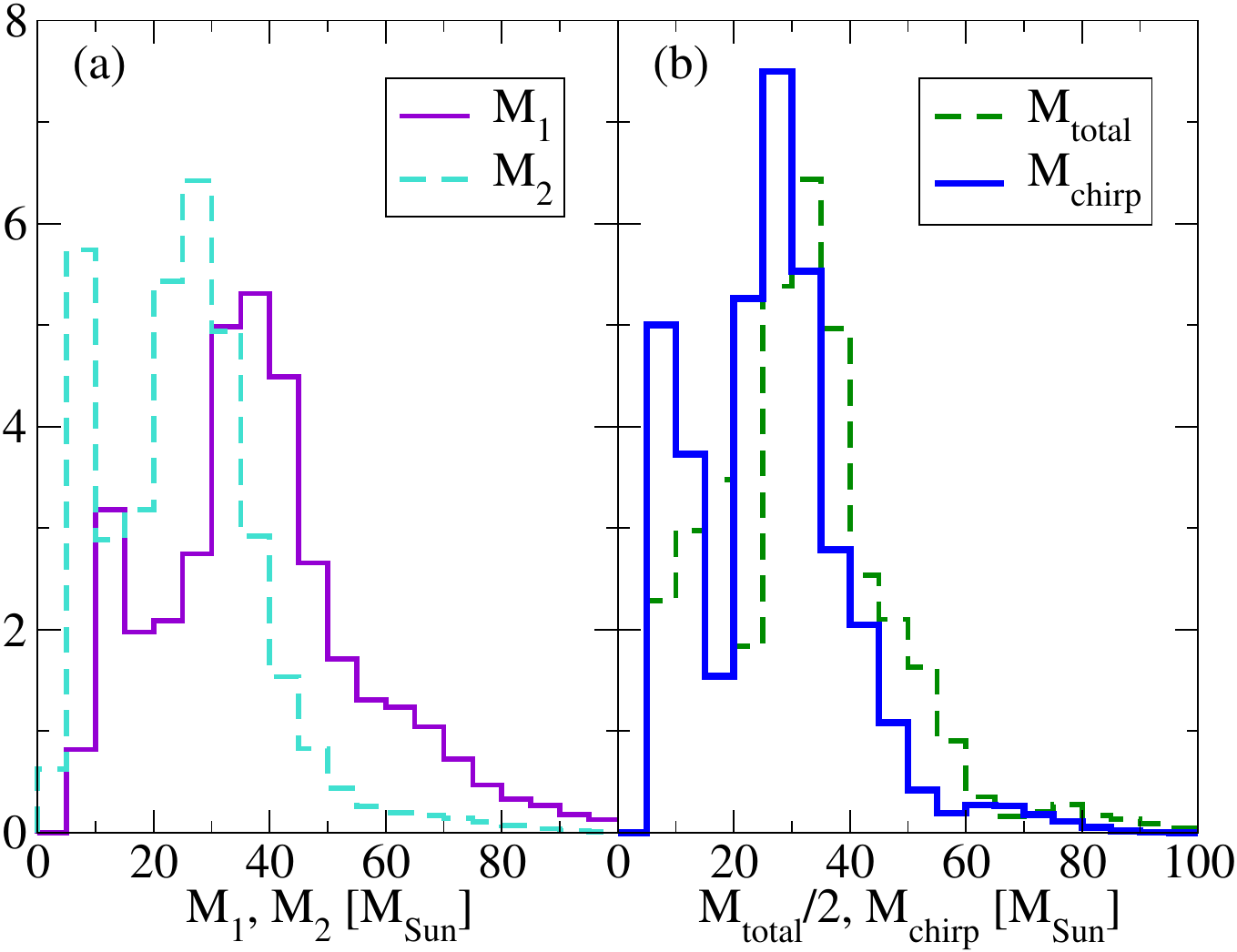}
    \caption{Probability number distributions of primary ($M_1$ in (a)), secondary ($M_2$ in (a)),
    total ($M_{\rm total}$ in (b)),
    and chirp ($\Mc$ in (b))
    masses of the observed 36 BBHs by using the GWTC-2 data~\citep{2020arXiv201014527A}.
    The lines correspond to the probability number of stars in the mass interval of $5\msun$.
    Note that the horizontal axis for the total mass is $\Mt/2$.
    }
    \label{fig:mass_distributions}
\end{figure}

\begin{figure}
    \centering
    \includegraphics[width=0.47\textwidth]{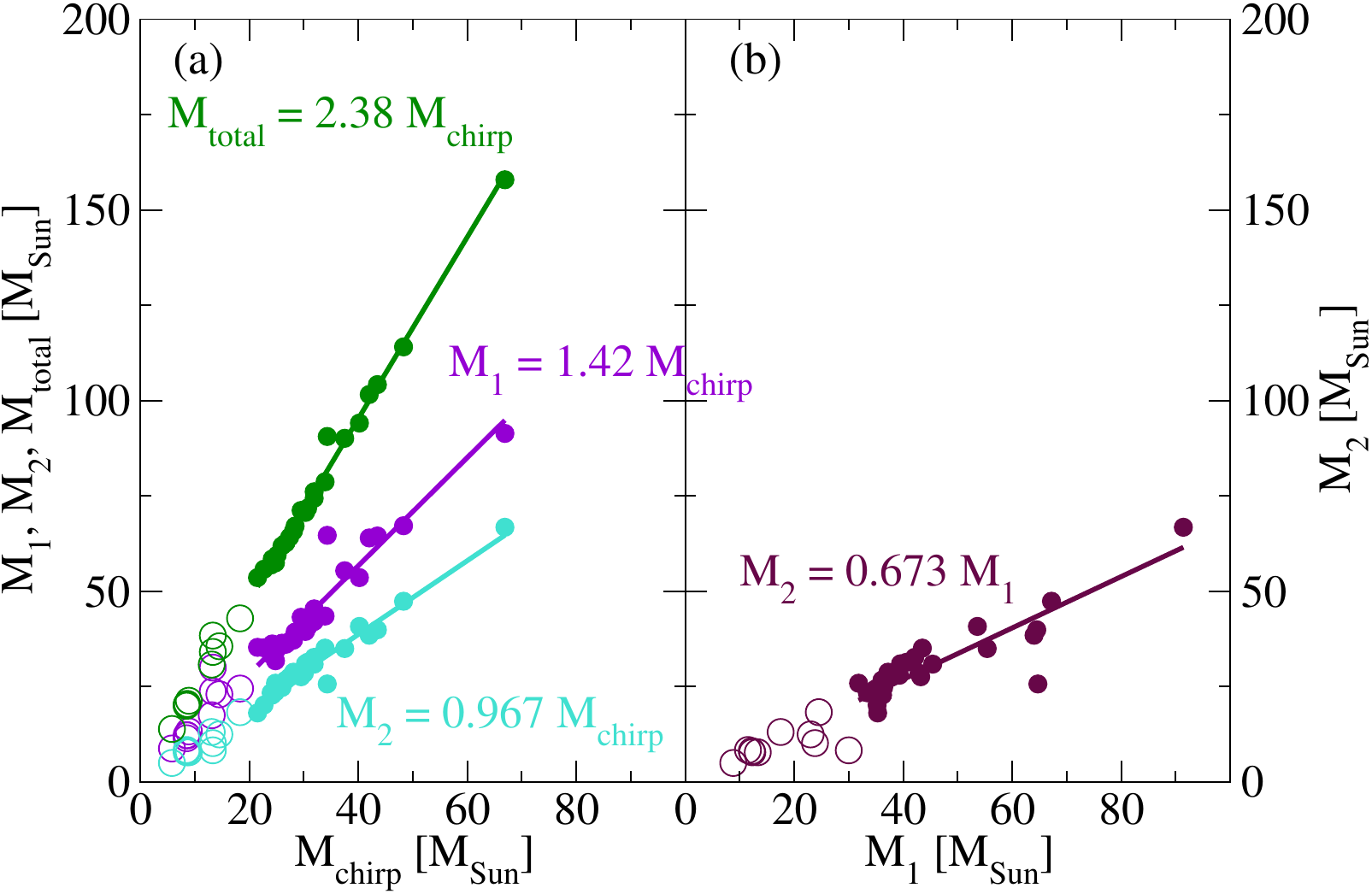}
    \caption{Distribution of median values of
    various masses; $M_1$ vs. $\Mc$, $M_2$ vs. $\Mc$,
    and $\Mt$ vs. $\Mc$ in (a), and $M_2$ vs. $M_1$ in (b).
    The circles and filled circles show BBHs
    with chirp mass
    lower or higher than $20\msun$, respectively.
    Each straight line shows the best fitting one for $\Mc > 20\msun$.}
    \label{fig:Mc_masses}
\end{figure}

The probability number distributions of primary ($M_1$), secondary ($M_2$), total ($M_{\rm total}$),
and chirp ($\Mc$) masses for the 36 BBHs
are shown in Figure~\ref{fig:mass_distributions} (a) and (b) where
the horizontal axis is shown in unit of $\msun$ 
while the lines correspond to the probability number of stars in the mass interval of $5\msun$. Note that we  use $\Mt/2$ for the total mass in Figure~\ref{fig:mass_distributions}  (b).
In all distributions, we can identify double peaks 
of probability number of stars. 
We confirmed that they exist even for
the case with the mass interval of $2.5\msun$.
The higher peaks of various mass are at $\sim 46\msun$,
$\sim 32\msun$, $\sim 77\msun$ and $\sim 32\msun$
for $M_1$, $M_2$, $\Mt$ and $\Mc$, respectively,
while the lower peaks  are at $\sim 18\msun$,
$\sim 10\msun$, $\sim 28\msun$ and $\sim 11\msun$
for $M_1$, $M_2$, $\Mt$ and $\Mc$, respectively.

In Figure~\ref{fig:Mc_masses} (a),
taking the median value of $\Mc$
as the horizontal axis, we plot the median value of $M_1$, $M_2$ and $\Mt$ for each event, respectively.
Here, the 10 circles and 26 filled circles show BBHs
with $\Mc$ lower and higher than $20\msun$, respectively.
We also show the median values of $M_2$ vs. $M_1$ in Figure~\ref{fig:Mc_masses} (b)
where we use the value of $\Mc$ but not the value of
$M_1$ and $M_2$ to determine if a certain point in the figure is a circle or a filled circle.
From Figure~\ref{fig:Mc_masses} (a) and Figure~\ref{fig:mass_distributions} (a) and (b), we can identify two groups of BBHs in the distributions of $M_1$, $M_2$ and $\Mt$ as a function of $\Mc$.
The boundary of two groups is at $\Mc \sim 20\msun$.
The lines in Figure~\ref{fig:Mc_masses}
(a) and (b) are the best linear fitting ones for $\Mc > 20\msun$. They are $M_1=1.42\Mc$, $M_2=0.967\Mc$, 
$\Mt =2.38\Mc$ and $M_2=0.673M_1$, respectively.
The correlation coefficients for each distribution are $0.957$, $0.978$, $0.996$ and $0.875$, respectively.
Since these are good correlations, 
there should be some physical explanations for them. 

We assume first that the correlation 
between $M_1$ and $M_2$ is given as $M_2=0.673 M_1$ from Figure~\ref{fig:Mc_masses} (b).
{From the definition of the chirp mass and $M_2=0.673 M_1$, 
we have $M_1=1.41\Mc$, $M_2=0.946\Mc$ and $\Mt=2.35\Mc$}
which agree quite well with $M_1=1.42\Mc$, $M_2=0.967\Mc$, 
$\Mt =2.38\Mc$ obtained from Figure~\ref{fig:Mc_masses} (a), respectively. This means that the relation of $M_2=0.673 M_1$ is the most important one, 
and thus other three relations are obtained from this $M_2=0.673 M_1$ relation.

What is the physical origin of the relation of 
$M_2=0.673 M_1$? 
We first notice that the smallest mass ratio $q~(=M_2/M_1)$
of 36 BBHs observed in O3a
is $\gtrsim 0.3$~\citep{2020arXiv201014533T} while 
in population synthesis models of Pop III stars such as in~\cite{Kinugawa2014},
the initial mass ratio $q$ of binaries ranges from $10\msun/M_1$ to $1$. Therefore, binaries 
with small mass ratio of $q < 0.3$ exist
when the binaries are first formed. 
However, the binaries with such small mass ratio tend to have large mass ratio due to mass transfer so that the fraction of merging Pop III BBH with $q \lesssim 0.5$ is much smaller than that of $q > 0.5$~\citep{Kinugawa2016c}.
In reality, most of the binary events in GWTC-2 satisfy
$0.6 < q < 0.8$ from Figure~\ref{fig:Mc_masses} (b).
Therefore, the relation of $M_2=0.673 M_1$ is consistent with Pop III star origin.

\begin{figure}
    \centering
    \includegraphics[width=\hsize,height=0.65\hsize,bb=90 5 630 422]{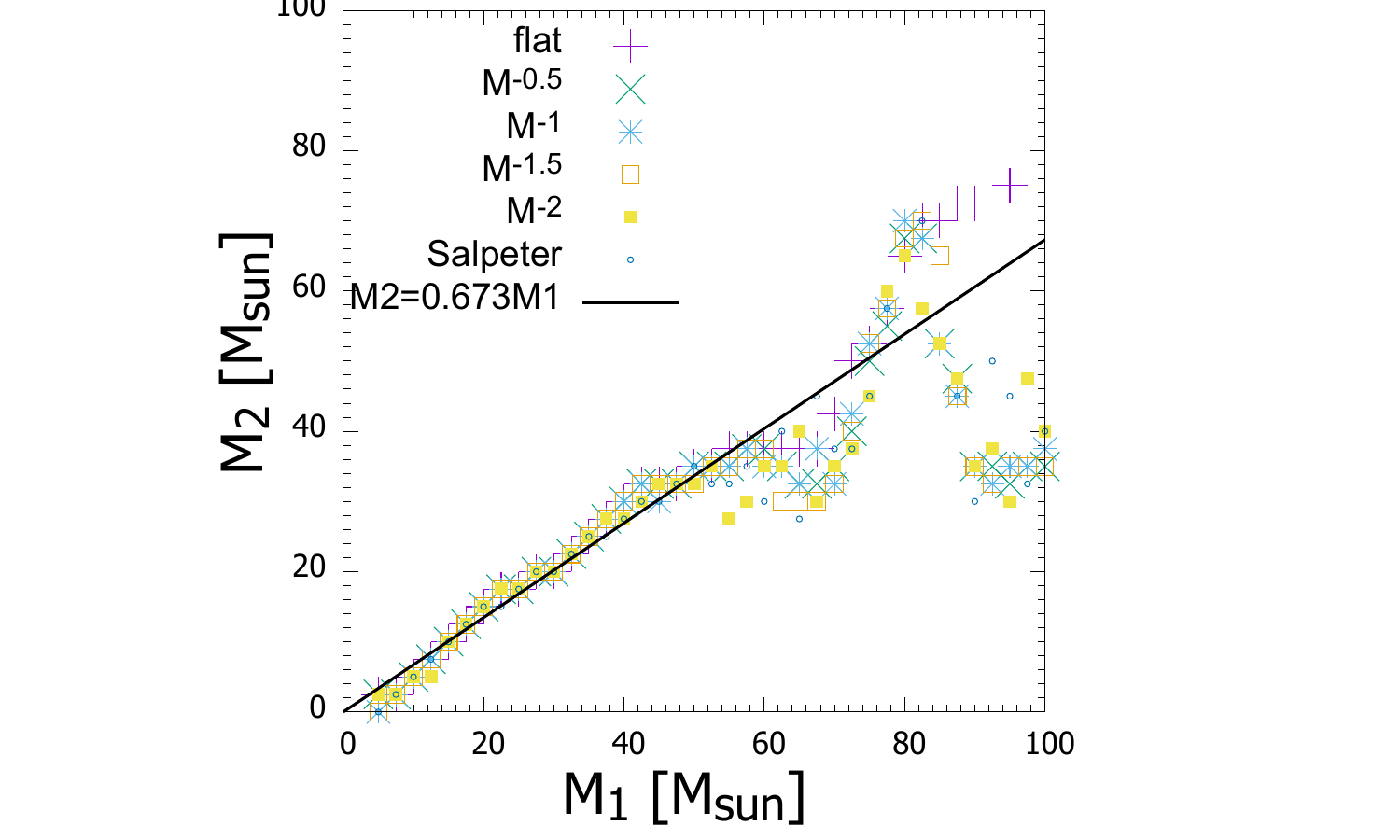}
    \caption{$M_2$ vs. $M_1$ of the BBHs which merge within the Hubble time, for various population synthesis simulation models of Pop III origin BBHs with IMF such as flat, $M^{-0.5}, M^{-1}, M^{-1.5}, M^{-2}$ and $M^{-2.35}$ (Salpeter), respectively. The mean values of median value of secondary mass $M_2$ for each primary mass $M_1$ are plotted. $M_2\cong 0.7M_1$ relation can be seen for all models up to $M_1\sim 50\msun$ similar to Figure~\ref{fig:Mc_masses} (b). For details of the simulations, see Section 2.2.}
    \label{fig:PopIII_m1m2}
\end{figure}

In Figure~\ref{fig:PopIII_m1m2}, the median values of secondary mass for each primary mass of Pop III BBHs which merge within the Hubble time are shown for various models. 
Although details of various population synthesis simulation models of Pop III BBHs  are described in the next section, one can identify $M_2\cong 0.7M_1$ relation similar to Figure~\ref{fig:Mc_masses} (b) for various models up to $M_1\sim 50\msun$.
{On the other hand, in the cases of Pop I/II field binaries and dynamical formation in dense star clusters, we estimate the median values of secondary mass using Figure 26 in \cite{Belczynski2020} and Figure 6 in \cite{Rodriguez:2019huv}. They follow the relations of $M_2\sim0.8M_1$, and $M_2\sim0.87M_1$, respectively.
The values of $M_2/M_1$ of these relations  are lager than those of the O3a observation and Pop III BBH simulations.
Especially, in the case of the dynamical formation, the mass ratio is much larger than those of Pop I/II and Pop III cases. The reason for this difference seems to come from repeated dynamical encounters, by which many exchanges of BHs make the binary to be a nearly equal-mass system although BBH was born with low mass ratio.}

\subsection{Results from population synthesis of Pop III binaries}

To calculate the number of events from the population synthesis simulations, we need the observable distance {(redshift)} for each binary.
Here, we treat only nonspinning equal-mass binaries
to calculate the maximum observable redshift
$z_{\rm max}$ for the LIGO O3a-Livingston (O3a-L)
by using the inspiral--merger--ringdown waveform shown
in~\cite{Nakamura:2016hna,2021arXiv210106402N} (see also \cite{Kinugawa2021massgap} for the O3a-L sensitivity curve).

\begin{figure}
    \centering
    \includegraphics[width=0.47\textwidth]{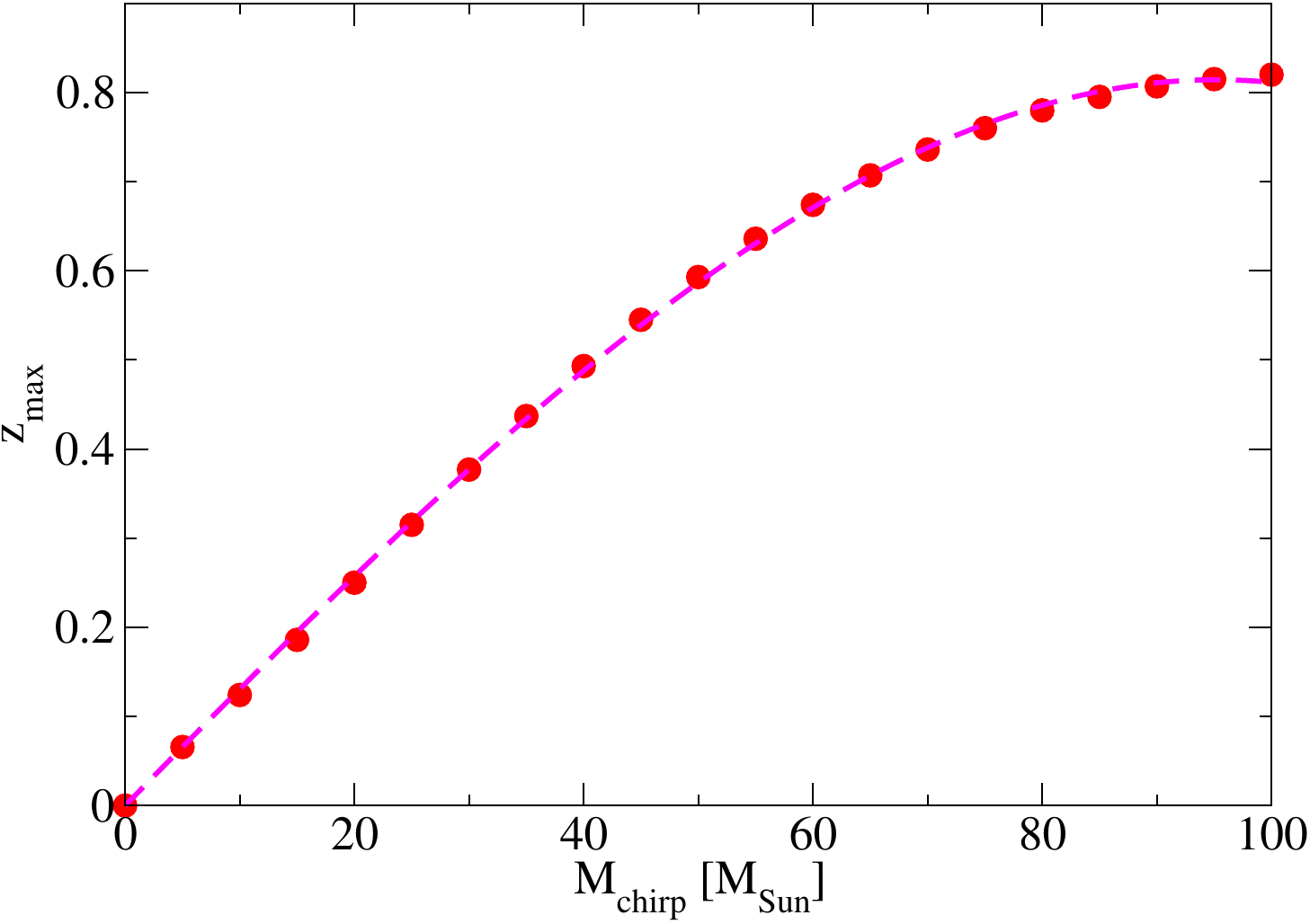}
    \caption{Maximum observable redshift $z_{\rm max}$ obtained by setting the averaged SNR $=8$ for nonspinning equal-mass BBHs with various $\Mc$
    for the LIGO O3a-Livingston (O3a-L) sensitivity.
    The dashed (magenta) curve is the fitting shown in Eq.~\eqref{eq:z_max}.}
    \label{fig:Mc_z}
\end{figure}

{For simplicity, we calculate the sky and polarization averaged SNR (Signal-to-Noise Ratio), and treat the chirp mass and redshift (where the luminosity distance, $D_L$ is a function of $z$) as the parameters of the waveform.
Although SNR also depends on the mass ratio,
we confirm for the O3a-L sensitivity that
the difference is at most $\sim 20\%$ in the estimation of luminosity distance
for binaries with redshifted chirp mass $(1+z)\Mc=5$--$126\msun$ in the case of $1/3 \leq q \leq 1$.
This means that we consider $\Mc=5$--$70\msun$ at $z \leq 0.8$.
For a BBH with a redshifted chirp mass, the SNR difference
is related to the difference in the luminosity distance directly, i.e., $D_L \propto 1/{\rm SNR}$.
This difference due to the assumption of equal-mass binaries is much smaller than errors ($\sim 300\%$)
in the LIGO/Virgo estimation of luminosity distances~\citep{2020arXiv201014527A}.
In the above assumption, the maximum observable redshift $z_{\rm max}$
for BBHs with various $\Mc$ by setting 
the averaged SNR $=8$ for O3a-L
is obtained as Figure~\ref{fig:Mc_z}}
and can be expressed by
a fitting function as
\begin{align}
z_{\rm max} =& 0.01319\,\frac{\Mc}{\msun}
-7.453 \times 10^{-6}\,\left(\frac{\Mc}{\msun}\right)^2
\cr &
-4.332 \times 10^{-7}\,\left(\frac{\Mc}{\msun}\right)^3 \,.
\label{eq:z_max}
\end{align}
{Note that when we apply the fitting functions shown in Figure~\ref{fig:Mc_masses} to evaluate the maximum observable redshift, the error in the maximum redshift due to the equal-mass assumption is only $\sim 3\%$.}

We simulate $10^6$ Pop III binary evolutions using the binary population synthesis code~\citep{Kinugawa2014,Kinugawa2020}.
In this Letter, $10$--$100\msun$ Pop III stellar evolutions
are discussed because of the following two reasons.
First, the typical mass of Pop III stars is considered
to be several tens solar
mass~\citep[e.g.][]{Hosokawa_2011,Hirano_2014,Susa_2014,Tarumi2020}.
Second, the observed BH masses with mass $\lesssim 80\msun$, can be explained by Pop III stars with initial mass 
$<100\msun$~\citep{Kinugawa2020}.

We treat 6 IMF models: flat, $M^{-0.5}$, $M^{-1}$, $M^{-1.5}$, $M^{-2}$, and Salpeter ($M^{-2.35}$) IMFs.
The flat model in this Letter is the same as M100 model
in~\cite{Kinugawa2020}.
In other models, we use the same initial distribution functions and binary parameters as those of the flat model except for IMF.
As for SFR as a function of the redshift $z$, we use that of~\cite{DeSouza_2011} with a factor of three smaller rate to be consistent with the restriction from CMB data 
by Planck~\citep{Visbal_2015,Inayoshi_2016}.  

Let us first define $R_{a,m}(t)$ as the merger rate density for individual mass `$m$' of the binary per co-moving volume at the cosmological time $t$ for a certain model `$a$' such as Pop III BBHs. Using Eq.~(90) in~\cite{Kinugawa2014},
$R_{a,m}(t)$ is given by
\begin{equation}
    R_{a,m}(t)=\int_0^t f_b\frac{{\rm SFR}(t')}
    {\langle M \rangle}\frac{N_{a,m}(t-t')}{N_{\rm tot}}dt'
    \,,
\label{eq:Ram}
\end{equation}
where $f_b$, ${\rm SFR}(t)$, $\langle M \rangle$, {$N_{a,m}(t-t')\,dt'$} and $N_{\rm tot}$ are the fraction of the total number of binaries to that of stars, 
SFR per co-moving volume per cosmic time,
mean mass of the stars, {the number of Pop III BBH merger events during $dt'$ for individual mass `$m$' with a delay time  $t-t'$} in a certain model `$a$',
and {the total number of Pop III binaries in the population synthesis simulation}, respectively. 
The physical meaning of the above equation is as follows.
First, because $f_b$={\rm (the number of binary stars)/(the number of stars)}, the maximum value of $f_b$ is $1/2$
while the fiducial value is $1/3$,
which means that $1/3$ of stars is a single star while $2/3$ is a binary.
Second, $\rm{SFR}(t')/\langle M \rangle$ is the formation  rate of the star in number.
We use the total mass density of Pop III stars is $\rho_{*,III}=6\times10^5~\msun\rm~Mpc^{-3}$~\citep{Inayoshi_2016}.
We assume that the redshift dependence of Pop III SFR is the  same as the Pop III SFR of~\cite{DeSouza_2011}. 
{Finally, $N_{a,m}(t-t')/N_{\rm tot}$ gives the fraction
of the Pop III BBH mergers with a delay time $t-t'$ for individual mass `$m$' 
in model `$a$'.} 


Using Eq.~(94) in~\cite{Kinugawa2014}, we obtain the expected number of events in  time interval of $\Delta t$ up to the redshift $z$ for individual mass `$m$' by 
{\begin{equation}
n_{a,m}(z) = 4\pi \int_0^z R_{a,m}(z') [r(z')]^2 
\frac{1}{1+z'} 
\frac{dr(z')}{dz'} dz' \times \Delta t
\,,
\label{eq:Nam}
\end{equation}
where $R_{a,m}(z)$ is given by using not `$t$' but `$z$' as an independent variable, $\Delta t=184/365.25$ yr is the observing time of O3a run assuming 100\% duty cycle, and $r(z)$ is the co-moving distance given by
$r(z) = (c/H_0) \int_0^z dz'
[\Omega_{\rm m}(1+z')^3+\Omega_{\rm \Lambda}]^{-1/2}$
where we use $H_0=67.74\, {\rm km\,s^{-1}\,Mpc^{-1}}$,
$\Omega_{\rm m}=0.3089$ and $\Omega_{\rm \Lambda}=1-\Omega_{\rm m}$ from~\cite{Planck_2015}.
In Eq.~\eqref{eq:Nam}, we set $z=z_{\rm max}$ evaluated by Eq.~\eqref{eq:z_max} with the chirp mass of each BBH under the assumption of equal mass binaries.} 

\begin{figure*}
\begin{tabular}{cc}
      \begin{minipage}[t]{0.48\hsize}
        \centering
\includegraphics[width=\hsize,height=0.65\hsize,bb=90 5 630 422]{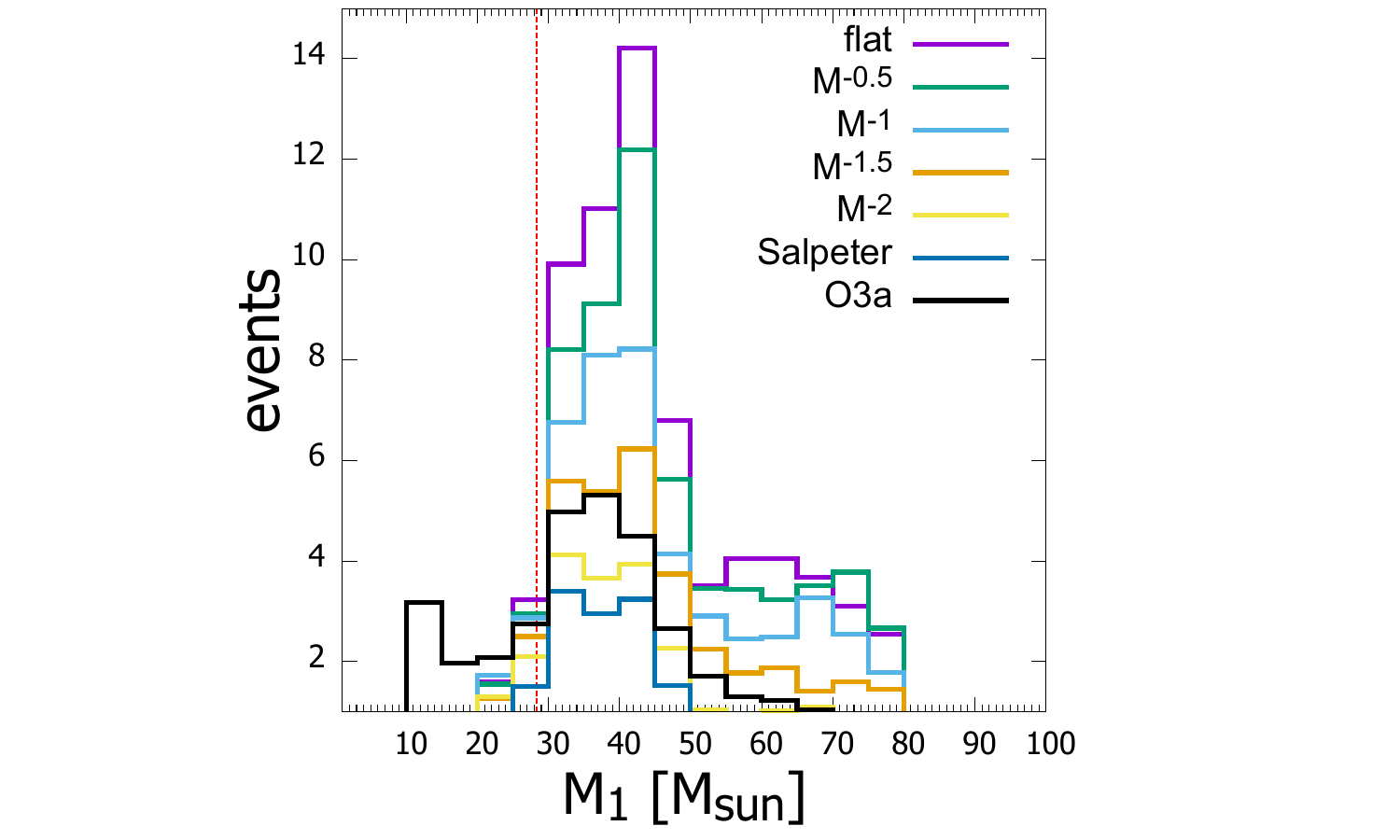}
\label{fig:ER_M1_PopIII}
\end{minipage} 
      \begin{minipage}[t]{0.48\hsize}
        \centering
\includegraphics[width=\hsize,height=0.65\hsize,bb=90 5 630 422]{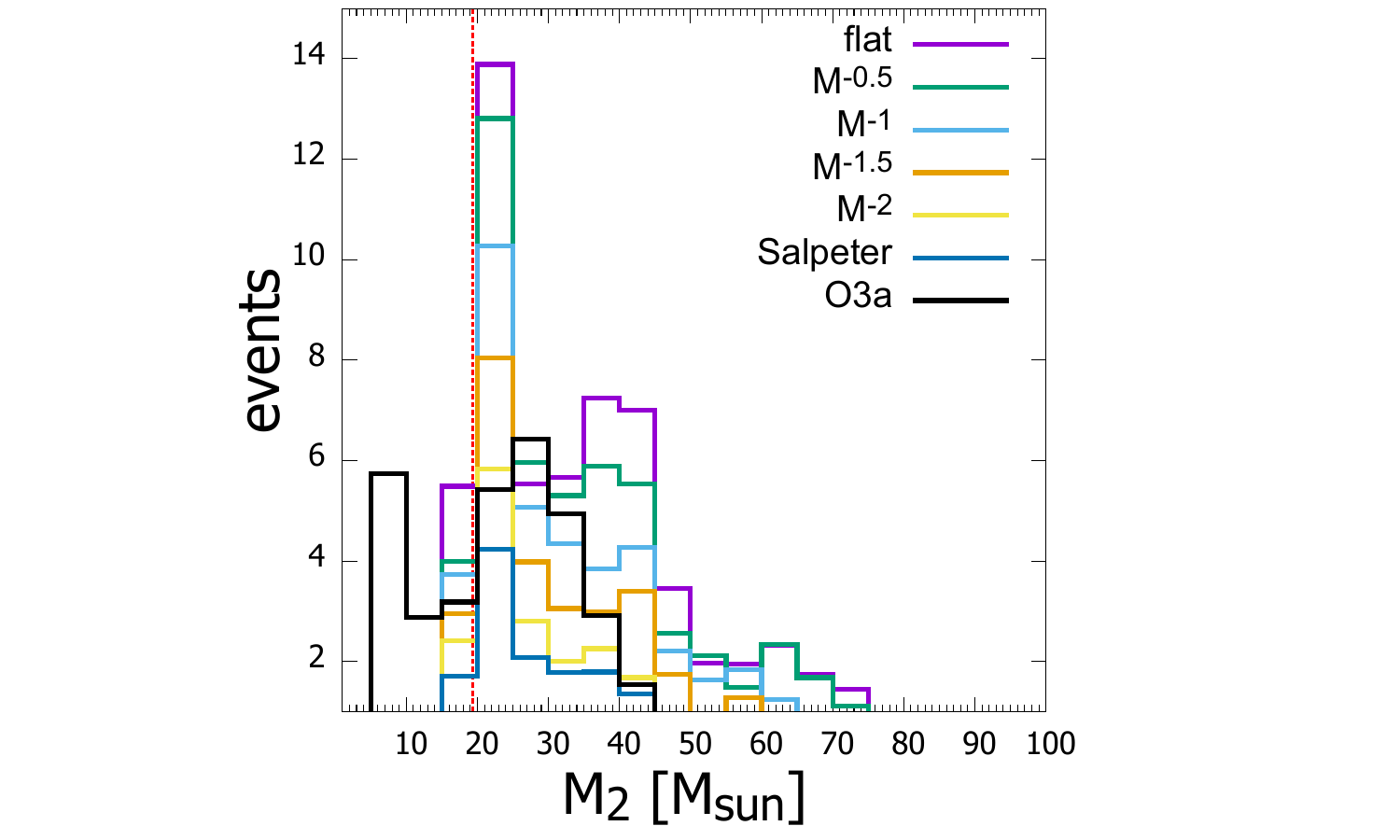}\\
\label{fig:ER_M2_PopIII}
\end{minipage} \\
      \begin{minipage}[t]{0.48\hsize}
        \centering
\includegraphics[width=\hsize,height=0.65\hsize,bb=90 5 630 422]{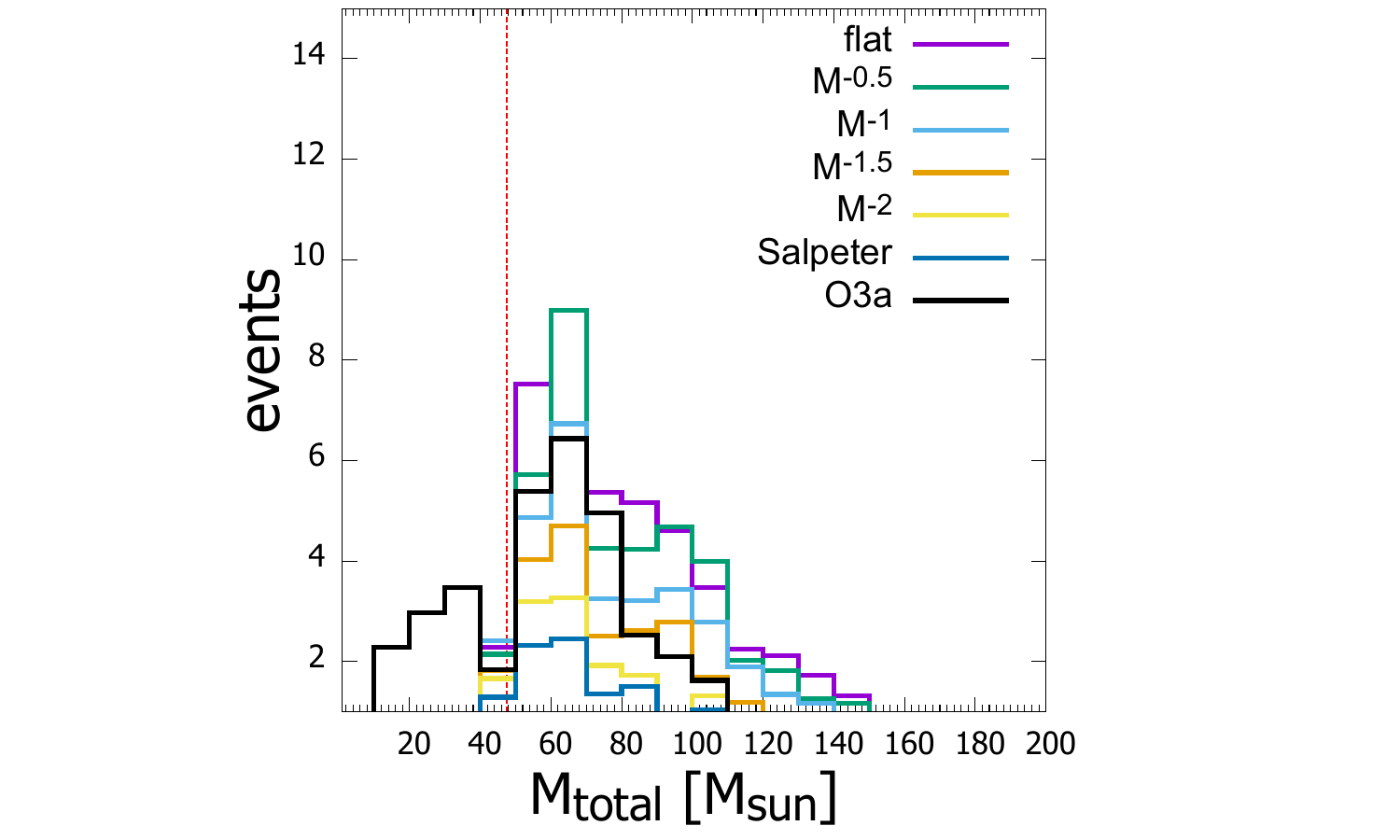}
\label{fig:ER_Mt_PopIII}
\end{minipage} 
      \begin{minipage}[t]{0.48\hsize}
        \centering
\includegraphics[width=\hsize,height=0.65\hsize,bb=90 5 630 422]{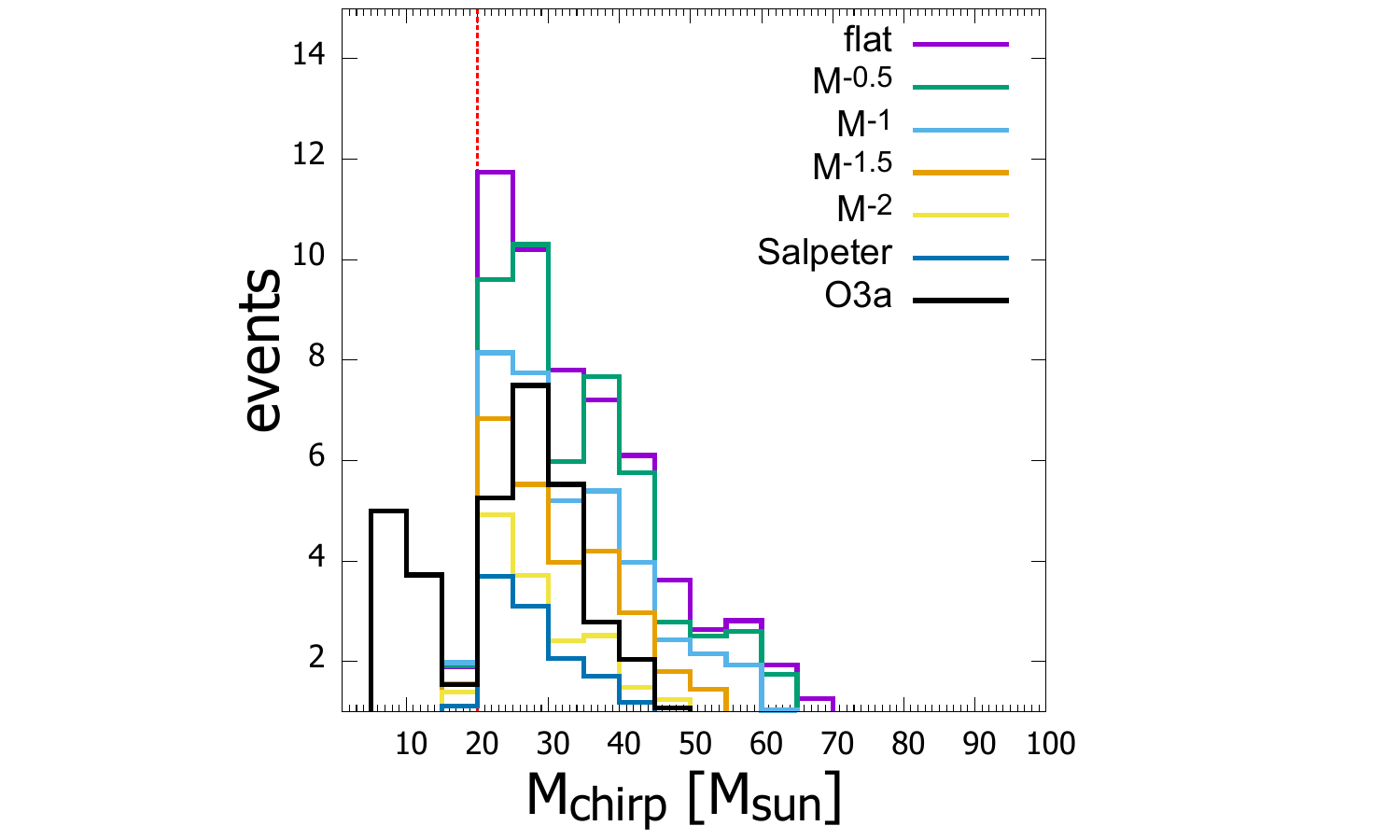}
\end{minipage} \\
    \end{tabular}
    \caption{ Four kinds of mass, ($M_1$, $M_2$, $\Mt$ and $\Mc$),  distribution of the number of events during the O3a observation time by the population synthesis simulations of Pop III BBHs using Eq.~\eqref{eq:Nam} for various theoretical models of IMF. The purple, green, blue, magenta, yellow and navy lines show the estimation for the flat, $M^{-0.5}$, $M^{-1}$, $M^{-1.5}$, $M^{-2}$ and Salpeter ($M^{-2.35}$) IMFs, respectively. 
    The black lines denote the O3a observation shown in Figure~\ref{fig:mass_distributions}.
    {The red vertical dotted lines correspond to $\Mc=20\msun$.}}
\label{fig:ER_Mc_PopIII}
\end{figure*}

\begin{table*}
\caption{Estimation of the minimum $\delta$ calculated
by using Eq.~\eqref{eq:est_delta} for $\Mc, M_1, M_2$ and $\Mt$
and the best fit value of the constant $\lambda$ shown in the parenthesis, respectively. We treat only the range of chirp mass of
$20\,\msun \leq \Mc \leq 100\,\msun$
and related ranges derived from the fitting functions shown in Figure~\ref{fig:Mc_masses} for the other masses. Here, $\lambda \leq 1.0$ since we assume $f_b=1/3$.
The boldface number shows the minimum value of $\delta$ for  $\Mc$, $M_1$, $M_2$ and $\Mt$, respectively. 
}
\label{tab:comparison}
\begin{center}
\begin{tabular}{lcccccc}
\hline
 Value of $\delta~(\lambda)$ & flat & $M^{-0.5}$ & $M^{-1}$ & $M^{-1.5}$ & $M^{-2}$ & Salpeter ($M^{-2.35}$) \\
\hline
$\Mc$ & 13.7 (0.518)& 14.72 (0.568) & {\bf 12.5} (0.734) & 13.8 (0.950) & 24.9 (1.00) & 35.8 (1.00) \\
$M_1$ & 6.54 (0.402) & 8.02 (0.465)& 4.38 (0.611) & {\bf 3.22} (0.821) & 5.45 (1.00) & 14.7 (1.00) \\
$M_2$ & 27.5 (0.463) & 23.0 (0.533) & {\bf 20.3} (0.678) & 20.7 (0.875) & 23.3 (1.00) & 33.9 (1.00) \\
$\Mt$ & {\bf 7.59} (0.654) & 10.9 ( 0.710) & 9.59 (0.919) & 12.1 (1.00) & 26.3 (1.00) & 42.0 (1.00) \\
\hline
\end{tabular}
\end{center}
\end{table*}

Figure~\ref{fig:ER_Mc_PopIII} shows four kinds of mass ($M_1$, $M_2$, $\Mt$ and $\Mc$) distribution of the number of events  during the O3a observation time by the population synthesis simulations of Pop III BBHs using Eq.~\eqref{eq:Nam} for various theoretical models of IMF.  {The red vertical dotted lines correspond to $\Mc=20\msun$.} The purple, green, blue, magenta, yellow and navy lines show the estimation for the flat, $M^{-0.5}$, $M^{-1}$, $M^{-1.5}$, $M^{-2}$ and Salpeter ($M^{-2.35}$) IMFs, respectively.
To determine which model is the best to explain the mass distributions of O3a data,
we compare the value of $\delta$ defined by
\begin{equation}
\label{eq:est_delta}
\delta = \sum_{i} (\lambda n_i^{\rm model}
-n_i^{\rm O3a})^2 \,,
\end{equation}
where $i$ is the number of each mass bin while
$n_i^{\rm O3a}$ and $n_i^{\rm model}$ are presented
in Figures~\ref{fig:mass_distributions}
and~\ref{fig:ER_Mc_PopIII}, respectively.
Using Eq.~\eqref{eq:est_delta},
we minimize $\delta$ as a function of $\lambda$.

In Eq.~\eqref{eq:est_delta},
we have introduced a variable $\lambda$
to fix SFR $\times f_b$ of the model. 
Here, assuming the fiducial value $f_b=1/3$, we restrict $\lambda \leq 1.0$.
In Table~\ref{tab:comparison}, we show estimation of the minimum $\delta$ calculated
by using Eq.~\eqref{eq:est_delta} for $\Mc$, $M_1$, $M_2$ and $\Mt$
and the best fit value of the constant $\lambda$ shown in the parenthesis, respectively. Here, $\lambda =0.734$, 
for example, means the model with 73.4\% of star formation rate compared with the fiducial one. We treat only the range of chirp mass of
$20\,\msun \leq \Mc \leq 100\,\msun$
and related ranges derived from the fitting functions shown in Figure~\ref{fig:Mc_masses} for the other mass.
The boldface number shows the minimum value of $\delta$ for  $\Mc$, $M_1$, $M_2$ and $\Mt$, respectively. 
Since the best IMF is not the same for $\Mc$, $M_1$, $M_2$ and $\Mt$, we can only state that $0\leq \alpha \leq 1.5$ is preferred from the O3a data.

\begin{figure}
    \centering
    \includegraphics[width=\hsize,height=0.65\hsize,bb=90 5 630 422]{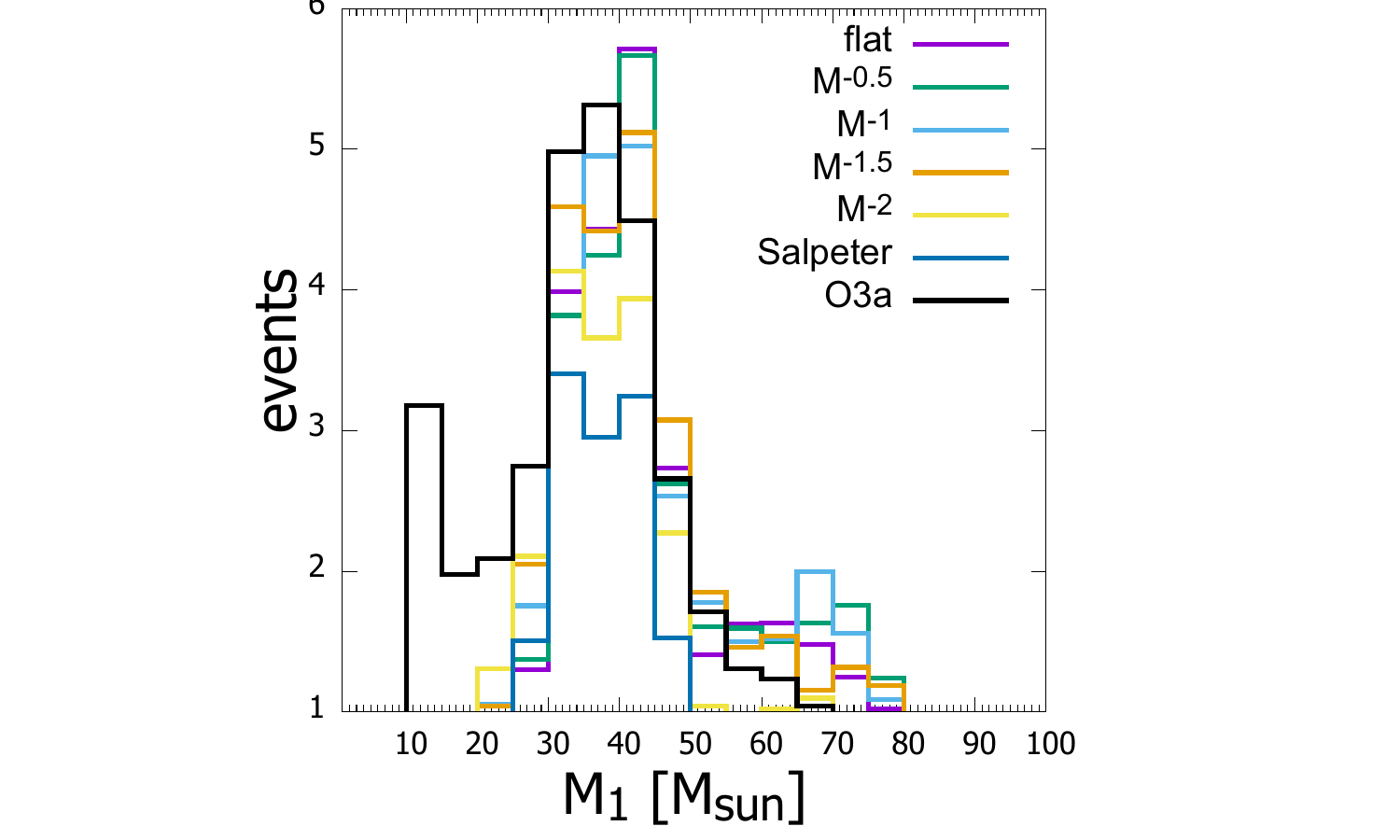}
    \caption{$M_1$ distribution for each IMF with the best value of $\lambda$ in Table~\ref{tab:comparison}.}
    \label{fig:M1_normalize}
\end{figure}

 {From Table~\ref{tab:comparison}, in the cases of $M_{chirp}$ and $M_2$,  the minima of $\delta$ are at $\alpha=1$.
On the other hand, in the case of $M_1$, there are two local minima at $\alpha=0$, and $1.5$, although the global minimum $\delta$ is at $\alpha=1.5$.
Furthermore, in the case of $M_{\rm total}$, there are also two local minima at $\alpha=0$, and $1$, although the global minimum of $\delta$ is at $\alpha=0$.
To make the situation here clearer, we show in Figure~\ref{fig:M1_normalize} the number distribution of $M_1$ with the minimized value of $\lambda$ for each $\alpha$ of IMF.
We see that although the fraction of massive BH slightly increases as $\alpha$ decreases, the mass distributions are almost the same for $0\leq \alpha \leq 1.5$.
The difference is too small to narrow down the candidates  using O3a events.}
We need more data than those of O3a in order to say which IMF is the best.
 {Our previous works \cite[e.g.][]{Kinugawa2016,Kinugawa2020} show that although the property of Pop III mass distribution hardly depends on binary parameters and initial distributions, the merger rate would change by a factor of a few.
Thus, $\lambda$ should depend on binary parameters and initial distributions.}

In conclusion, our population synthesis simulations of Pop III stars show that for $\Mc > 20\msun$, four kinds of BH masses, i.e., $M_1$, $M_2$, $\Mt$ and $\Mc$, distributions of O3a data are consistent with a moderately decreasing IMF in the form of $f(M) \propto M^{-\alpha}$ with $0\leq \alpha \leq 1.5$. The distribution of $M_1$ and $M_2$ of simulation data agrees with $M_2\cong 0.7M_1$ relation from O3a. 

\section*{Acknowledgment}

T. K. acknowledges support from the University of Tokyo Young Excellent Researcher program.
H. N. acknowledges support from
JSPS KAKENHI Grant Nos. JP16K05347 and JP17H06358.

\section*{Data Availability}

Results will be shared on reasonable request to corresponding author.

\bibliographystyle{mnras}

\bibliography{ref}

\begin{thebibliography}{}
\makeatletter
\relax
\def\mn@urlcharsother{\let\do\@makeother \do\$\do\&\do\#\do\^\do\_\do\%\do\~}
\def\mn@doi{\begingroup\mn@urlcharsother \@ifnextchar [ {\mn@doi@}
  {\mn@doi@[]}}
\def\mn@doi@[#1]#2{\def\@tempa{#1}\ifx\@tempa\@empty \href
  {http://dx.doi.org/#2} {doi:#2}\else \href {http://dx.doi.org/#2} {#1}\fi
  \endgroup}
\def\mn@eprint#1#2{\mn@eprint@#1:#2::\@nil}
\def\mn@eprint@arXiv#1{\href {http://arxiv.org/abs/#1} {{\tt arXiv:#1}}}
\def\mn@eprint@dblp#1{\href {http://dblp.uni-trier.de/rec/bibtex/#1.xml}
  {dblp:#1}}
\def\mn@eprint@#1:#2:#3:#4\@nil{\def\@tempa {#1}\def\@tempb {#2}\def\@tempc
  {#3}\ifx \@tempc \@empty \let \@tempc \@tempb \let \@tempb \@tempa \fi \ifx
  \@tempb \@empty \def\@tempb {arXiv}\fi \@ifundefined
  {mn@eprint@\@tempb}{\@tempb:\@tempc}{\expandafter \expandafter \csname
  mn@eprint@\@tempb\endcsname \expandafter{\@tempc}}}

\bibitem[\protect\citeauthoryear{{Abbott} et~al.,}{{Abbott}
  et~al.}{2019}]{2019PhRvX...9c1040A}
{Abbott} B.~P.,  et~al., 2019, \mn@doi [Physical Review X]
  {10.1103/PhysRevX.9.031040}, \href
  {https://ui.adsabs.harvard.edu/abs/2019PhRvX...9c1040A} {9, 031040}

\bibitem[\protect\citeauthoryear{{Abbott} et~al.,}{{Abbott}
  et~al.}{2020}]{2020arXiv201014527A}
{Abbott} R.,  et~al., 2020, arXiv e-prints, \href
  {https://ui.adsabs.harvard.edu/abs/2020arXiv201014527A} {p. arXiv:2010.14527}

\bibitem[\protect\citeauthoryear{{Antonini} \& {Gieles}}{{Antonini} \&
  {Gieles}}{2020}]{2020PhRvD.102l3016A}
{Antonini} F.,  {Gieles} M.,  2020, \mn@doi [\prd]
  {10.1103/PhysRevD.102.123016}, \href
  {https://ui.adsabs.harvard.edu/abs/2020PhRvD.102l3016A} {102, 123016}

\bibitem[\protect\citeauthoryear{{Banerjee}}{{Banerjee}}{2020a}]{2020arXiv201107000B}
{Banerjee} S.,  2020a, arXiv e-prints, \href
  {https://ui.adsabs.harvard.edu/abs/2020arXiv201107000B} {p. arXiv:2011.07000}

\bibitem[\protect\citeauthoryear{{Banerjee}}{{Banerjee}}{2020b}]{2020MNRAS.500.3002B}
{Banerjee} S.,  2020b, \mn@doi [\mnras] {10.1093/mnras/staa2392}, \href
  {https://ui.adsabs.harvard.edu/abs/2020MNRAS.500.3002B} {500, 3002}

\bibitem[\protect\citeauthoryear{{Belczynski} et~al.,}{{Belczynski}
  et~al.}{2020}]{Belczynski2020}
{Belczynski} K.,  et~al., 2020, \mn@doi [\aap] {10.1051/0004-6361/201936528},
  \href {https://ui.adsabs.harvard.edu/abs/2020A&A...636A.104B} {636, A104}

\bibitem[\protect\citeauthoryear{{Bouffanais}, {Mapelli}, {Santoliquido},
  {Giacobbo}, {Di Carlo}, {Rastello}, {Artale}  \& {Iorio}}{{Bouffanais}
  et~al.}{2021}]{2021arXiv210212495B}
{Bouffanais} Y.,  {Mapelli} M.,  {Santoliquido} F.,  {Giacobbo} N.,  {Di Carlo}
  U.~N.,  {Rastello} S.,  {Artale} M.~C.,   {Iorio} G.,  2021, arXiv e-prints,
  \href {https://ui.adsabs.harvard.edu/abs/2021arXiv210212495B} {p.
  arXiv:2102.12495}

\bibitem[\protect\citeauthoryear{{Callister}, {Farr}  \& {Renzo}}{{Callister}
  et~al.}{2020}]{2020arXiv201109570C}
{Callister} T.~A.,  {Farr} W.~M.,   {Renzo} M.,  2020, arXiv e-prints, \href
  {https://ui.adsabs.harvard.edu/abs/2020arXiv201109570C} {p. arXiv:2011.09570}

\bibitem[\protect\citeauthoryear{{De Luca}, {Franciolini}, {Pani}  \&
  {Riotto}}{{De Luca} et~al.}{2021}]{2021arXiv210203809D}
{De Luca} V.,  {Franciolini} G.,  {Pani} P.,   {Riotto} A.,  2021, arXiv
  e-prints, \href {https://ui.adsabs.harvard.edu/abs/2021arXiv210203809D} {p.
  arXiv:2102.03809}

\bibitem[\protect\citeauthoryear{{Deng}}{{Deng}}{2021}]{2021arXiv210111098D}
{Deng} H.,  2021, arXiv e-prints, \href
  {https://ui.adsabs.harvard.edu/abs/2021arXiv210111098D} {p. arXiv:2101.11098}

\bibitem[\protect\citeauthoryear{{Farrell}, {Groh}, {Hirschi}, {Murphy},
  {Kaiser}, {Ekstr{\"o}m}, {Georgy}  \& {Meynet}}{{Farrell}
  et~al.}{2020}]{Farrell2020}
{Farrell} E.~J.,  {Groh} J.~H.,  {Hirschi} R.,  {Murphy} L.,  {Kaiser} E.,
  {Ekstr{\"o}m} S.,  {Georgy} C.,   {Meynet} G.,  2020, arXiv e-prints, \href
  {https://ui.adsabs.harvard.edu/abs/2020arXiv200906585F} {p. arXiv:2009.06585}

\bibitem[\protect\citeauthoryear{{Fishbach} et~al.,}{{Fishbach}
  et~al.}{2021}]{2021arXiv210107699F}
{Fishbach} M.,  et~al., 2021, arXiv e-prints, \href
  {https://ui.adsabs.harvard.edu/abs/2021arXiv210107699F} {p. arXiv:2101.07699}

\bibitem[\protect\citeauthoryear{{Fragione} \& {Loeb}}{{Fragione} \&
  {Loeb}}{2021}]{2021MNRAS.tmp..279F}
{Fragione} G.,  {Loeb} A.,  2021, \mn@doi [\mnras] {10.1093/mnras/stab247},
  \href {https://ui.adsabs.harvard.edu/abs/2021MNRAS.tmp..279F} {}

\bibitem[\protect\citeauthoryear{{Gerosa}, {Mould}, {Gangardt}, {Schmidt},
  {Pratten}  \& {Thomas}}{{Gerosa} et~al.}{2020}]{2020arXiv201111948G}
{Gerosa} D.,  {Mould} M.,  {Gangardt} D.,  {Schmidt} P.,  {Pratten} G.,
  {Thomas} L.~M.,  2020, arXiv e-prints, \href
  {https://ui.adsabs.harvard.edu/abs/2020arXiv201111948G} {p. arXiv:2011.11948}

\bibitem[\protect\citeauthoryear{Hall, Gow  \& Byrnes}{Hall
  et~al.}{2020}]{Hall:2020daa}
Hall A.,  Gow A.~D.,   Byrnes C.~T.,  2020, \mn@doi [Phys. Rev. D]
  {10.1103/PhysRevD.102.123524}, 102, 123524

\bibitem[\protect\citeauthoryear{{Hirano}, {Hosokawa}, {Yoshida}, {Umeda},
  {Omukai}, {Chiaki}  \& {Yorke}}{{Hirano} et~al.}{2014}]{Hirano_2014}
{Hirano} S.,  {Hosokawa} T.,  {Yoshida} N.,  {Umeda} H.,  {Omukai} K.,
  {Chiaki} G.,   {Yorke} H.~W.,  2014, \mn@doi [\apj]
  {10.1088/0004-637X/781/2/60}, \href
  {http://adsabs.harvard.edu/abs/2014ApJ...781...60H} {781, 60}

\bibitem[\protect\citeauthoryear{{Hosokawa}, {Omukai}, {Yoshida}  \&
  {Yorke}}{{Hosokawa} et~al.}{2011}]{Hosokawa_2011}
{Hosokawa} T.,  {Omukai} K.,  {Yoshida} N.,   {Yorke} H.~W.,  2011, \mn@doi
  [Science] {10.1126/science.1207433}, \href
  {http://adsabs.harvard.edu/abs/2011Sci...334.1250H} {334, 1250}

\bibitem[\protect\citeauthoryear{{H{\"u}tsi}, {Raidal}, {Vaskonen}  \&
  {Veerm{\"a}e}}{{H{\"u}tsi} et~al.}{2020}]{2020arXiv201202786H}
{H{\"u}tsi} G.,  {Raidal} M.,  {Vaskonen} V.,   {Veerm{\"a}e} H.,  2020, arXiv
  e-prints, \href {https://ui.adsabs.harvard.edu/abs/2020arXiv201202786H} {p.
  arXiv:2012.02786}

\bibitem[\protect\citeauthoryear{{Inayoshi}, {Kashiyama}, {Visbal}  \&
  {Haiman}}{{Inayoshi} et~al.}{2016}]{Inayoshi_2016}
{Inayoshi} K.,  {Kashiyama} K.,  {Visbal} E.,   {Haiman} Z.,  2016, \mn@doi
  [\mnras] {10.1093/mnras/stw1431}, \href
  {http://adsabs.harvard.edu/abs/2016MNRAS.461.2722I} {461, 2722}

\bibitem[\protect\citeauthoryear{{Kimball} et~al.,}{{Kimball}
  et~al.}{2020a}]{2020arXiv201105332K}
{Kimball} C.,  et~al., 2020a, arXiv e-prints, \href
  {https://ui.adsabs.harvard.edu/abs/2020arXiv201105332K} {p. arXiv:2011.05332}

\bibitem[\protect\citeauthoryear{{Kimball}, {Talbot}, {Berry}, {Carney},
  {Zevin}, {Thrane}  \& {Kalogera}}{{Kimball}
  et~al.}{2020b}]{2020ApJ...900..177K}
{Kimball} C.,  {Talbot} C.,  {Berry} C. P.~L.,  {Carney} M.,  {Zevin} M.,
  {Thrane} E.,   {Kalogera} V.,  2020b, \mn@doi [\apj]
  {10.3847/1538-4357/aba518}, \href
  {https://ui.adsabs.harvard.edu/abs/2020ApJ...900..177K} {900, 177}

\bibitem[\protect\citeauthoryear{{Kinugawa}, {Inayoshi}, {Hotokezaka},
  {Nakauchi}  \& {Nakamura}}{{Kinugawa} et~al.}{2014}]{Kinugawa2014}
{Kinugawa} T.,  {Inayoshi} K.,  {Hotokezaka} K.,  {Nakauchi} D.,   {Nakamura}
  T.,  2014, \mn@doi [\mnras] {10.1093/mnras/stu1022}, \href
  {http://adsabs.harvard.edu/abs/2014MNRAS.442.2963K} {442, 2963}

\bibitem[\protect\citeauthoryear{{Kinugawa}, {Miyamoto}, {Kanda}  \&
  {Nakamura}}{{Kinugawa} et~al.}{2016a}]{Kinugawa2016}
{Kinugawa} T.,  {Miyamoto} A.,  {Kanda} N.,   {Nakamura} T.,  2016a, \mn@doi
  [\mnras] {10.1093/mnras/stv2624}, \href
  {http://adsabs.harvard.edu/abs/2016MNRAS.456.1093K} {456, 1093}

\bibitem[\protect\citeauthoryear{{Kinugawa}, {Nakano}  \&
  {Nakamura}}{{Kinugawa} et~al.}{2016b}]{Kinugawa2016c}
{Kinugawa} T.,  {Nakano} H.,   {Nakamura} T.,  2016b, \mn@doi [Progress of
  Theoretical and Experimental Physics] {10.1093/ptep/ptw143}, \href
  {https://ui.adsabs.harvard.edu/abs/2016PTEP.2016j3E01K} {2016, 103E01}

\bibitem[\protect\citeauthoryear{{Kinugawa}, {Nakamura}  \&
  {Nakano}}{{Kinugawa} et~al.}{2020}]{Kinugawa2020}
{Kinugawa} T.,  {Nakamura} T.,   {Nakano} H.,  2020, \mn@doi [\mnras]
  {10.1093/mnras/staa2511}, \href
  {https://ui.adsabs.harvard.edu/abs/2020MNRAS.498.3946K} {498, 3946}

\bibitem[\protect\citeauthoryear{{Kinugawa}, {Nakamura}  \&
  {Nakano}}{{Kinugawa} et~al.}{2021a}]{Kinugawa2021}
{Kinugawa} T.,  {Nakamura} T.,   {Nakano} H.,  2021a, \mn@doi [\mnras]
  {10.1093/mnrasl/slaa191}, \href
  {https://ui.adsabs.harvard.edu/abs/2021MNRAS.501L..49K} {501, L49}

\bibitem[\protect\citeauthoryear{{Kinugawa}, {Nakamura}  \&
  {Nakano}}{{Kinugawa} et~al.}{2021b}]{Kinugawa2021massgap}
{Kinugawa} T.,  {Nakamura} T.,   {Nakano} H.,  2021b, \mn@doi [Progress of
  Theoretical and Experimental Physics] {10.1093/ptep/ptaa176}, \href
  {https://ui.adsabs.harvard.edu/abs/2021PTEP.2021b1E01K} {2021, 021E01}

\bibitem[\protect\citeauthoryear{Nakamura et~al.}{Nakamura
  et~al.}{2016}]{Nakamura:2016hna}
Nakamura T.,  et~al., 2016, \mn@doi [PTEP] {10.1093/ptep/ptw127}, 2016, 093E01

\bibitem[\protect\citeauthoryear{{Nakano}, {Fujita}, {Isoyama}  \&
  {Sago}}{{Nakano} et~al.}{2021}]{2021arXiv210106402N}
{Nakano} H.,  {Fujita} R.,  {Isoyama} S.,   {Sago} N.,  2021, arXiv e-prints,
  \href {https://ui.adsabs.harvard.edu/abs/2021arXiv210106402N} {p.
  arXiv:2101.06402}

\bibitem[\protect\citeauthoryear{{Planck Collaboration} et~al.,}{{Planck
  Collaboration} et~al.}{2016}]{Planck_2015}
{Planck Collaboration} et~al., 2016, \mn@doi [\aap]
  {10.1051/0004-6361/201525830}, \href
  {https://ui.adsabs.harvard.edu/abs/2016A&A...594A..13P} {594, A13}

\bibitem[\protect\citeauthoryear{Rodriguez, Amaro-Seoane, Chatterjee  \&
  Rasio}{Rodriguez et~al.}{2018}]{Rodriguez:2017pec}
Rodriguez C.~L.,  Amaro-Seoane P.,  Chatterjee S.,   Rasio F.~A.,  2018,
  \mn@doi [Phys. Rev. Lett.] {10.1103/PhysRevLett.120.151101}, 120, 151101

\bibitem[\protect\citeauthoryear{Rodriguez, Zevin, Amaro-Seoane, Chatterjee,
  Kremer, Rasio  \& Ye}{Rodriguez et~al.}{2019}]{Rodriguez:2019huv}
Rodriguez C.~L.,  Zevin M.,  Amaro-Seoane P.,  Chatterjee S.,  Kremer K.,
  Rasio F.~A.,   Ye C.~S.,  2019, \mn@doi [Phys. Rev. D]
  {10.1103/PhysRevD.100.043027}, 100, 043027

\bibitem[\protect\citeauthoryear{{Rodriguez}, {Kremer}, {Chatterjee},
  {Fragione}, {Loeb}, {Rasio}, {Weatherford}  \& {Ye}}{{Rodriguez}
  et~al.}{2021}]{2021arXiv210107793R}
{Rodriguez} C.~L.,  {Kremer} K.,  {Chatterjee} S.,  {Fragione} G.,  {Loeb} A.,
  {Rasio} F.~A.,  {Weatherford} N.~C.,   {Ye} C.~S.,  2021, arXiv e-prints,
  \href {https://ui.adsabs.harvard.edu/abs/2021arXiv210107793R} {p.
  arXiv:2101.07793}

\bibitem[\protect\citeauthoryear{{Susa}, {Hasegawa}  \& {Tominaga}}{{Susa}
  et~al.}{2014}]{Susa_2014}
{Susa} H.,  {Hasegawa} K.,   {Tominaga} N.,  2014, \mn@doi [\apj]
  {10.1088/0004-637X/792/1/32}, \href
  {https://ui.adsabs.harvard.edu/abs/2014ApJ...792...32S} {792, 32}

\bibitem[\protect\citeauthoryear{{Tanikawa}, {Susa}, {Yoshida}, {Trani}  \&
  {Kinugawa}}{{Tanikawa} et~al.}{2020}]{Tanikawa2020}
{Tanikawa} A.,  {Susa} H.,  {Yoshida} T.,  {Trani} A.~A.,   {Kinugawa} T.,
  2020, arXiv e-prints, \href
  {https://ui.adsabs.harvard.edu/abs/2020arXiv200801890T} {p. arXiv:2008.01890}

\bibitem[\protect\citeauthoryear{{Tarumi}, {Hartwig}  \& {Magg}}{{Tarumi}
  et~al.}{2020}]{Tarumi2020}
{Tarumi} Y.,  {Hartwig} T.,   {Magg} M.,  2020, \mn@doi [\apj]
  {10.3847/1538-4357/ab960d}, \href
  {https://ui.adsabs.harvard.edu/abs/2020ApJ...897...58T} {897, 58}

\bibitem[\protect\citeauthoryear{{The LIGO Scientific Collaboration}
  et~al.,}{{The LIGO Scientific Collaboration}
  et~al.}{2020}]{2020arXiv201014533T}
{The LIGO Scientific Collaboration} et~al., 2020, arXiv e-prints, \href
  {https://ui.adsabs.harvard.edu/abs/2020arXiv201014533T} {p. arXiv:2010.14533}

\bibitem[\protect\citeauthoryear{{Tiwari}}{{Tiwari}}{2020}]{2020arXiv200615047T}
{Tiwari} V.,  2020, arXiv e-prints, \href
  {https://ui.adsabs.harvard.edu/abs/2020arXiv200615047T} {p. arXiv:2006.15047}

\bibitem[\protect\citeauthoryear{{Tiwari} \& {Fairhurst}}{{Tiwari} \&
  {Fairhurst}}{2020}]{2020arXiv201104502T}
{Tiwari} V.,  {Fairhurst} S.,  2020, arXiv e-prints, \href
  {https://ui.adsabs.harvard.edu/abs/2020arXiv201104502T} {p. arXiv:2011.04502}

\bibitem[\protect\citeauthoryear{{Trani}, {Tanikawa}, {Fujii}, {Leigh}  \&
  {Kumamoto}}{{Trani} et~al.}{2021}]{2021arXiv210201689T}
{Trani} A.~A.,  {Tanikawa} A.,  {Fujii} M.~S.,  {Leigh} N. W.~C.,   {Kumamoto}
  J.,  2021, arXiv e-prints, \href
  {https://ui.adsabs.harvard.edu/abs/2021arXiv210201689T} {p. arXiv:2102.01689}

\bibitem[\protect\citeauthoryear{{Veske}, {Sullivan}, {M{\'a}rka}, {Bartos},
  {Corley}, {Samsing}, {Buscicchio}  \& {M{\'a}rka}}{{Veske}
  et~al.}{2021}]{2021ApJ...907L..48V}
{Veske} D.,  {Sullivan} A.~G.,  {M{\'a}rka} Z.,  {Bartos} I.,  {Corley} K.~R.,
  {Samsing} J.,  {Buscicchio} R.,   {M{\'a}rka} S.,  2021, \mn@doi [\apjl]
  {10.3847/2041-8213/abd721}, \href
  {https://ui.adsabs.harvard.edu/abs/2021ApJ...907L..48V} {907, L48}

\bibitem[\protect\citeauthoryear{{Visbal}, {Haiman}  \& {Bryan}}{{Visbal}
  et~al.}{2015}]{Visbal_2015}
{Visbal} E.,  {Haiman} Z.,   {Bryan} G.~L.,  2015, \mn@doi [\mnras]
  {10.1093/mnras/stv1941}, \href
  {http://adsabs.harvard.edu/abs/2015MNRAS.453.4456V} {453, 4456}

\bibitem[\protect\citeauthoryear{{Wong}, {Franciolini}, {De Luca}, {Baibhav},
  {Berti}, {Pani}  \& {Riotto}}{{Wong} et~al.}{2020a}]{2020arXiv201101865W}
{Wong} K.~W.~K.,  {Franciolini} G.,  {De Luca} V.,  {Baibhav} V.,  {Berti} E.,
  {Pani} P.,   {Riotto} A.,  2020a, arXiv e-prints, \href
  {https://ui.adsabs.harvard.edu/abs/2020arXiv201101865W} {p. arXiv:2011.01865}

\bibitem[\protect\citeauthoryear{{Wong}, {Breivik}, {Kremer}  \&
  {Callister}}{{Wong} et~al.}{2020b}]{2020arXiv201103564W}
{Wong} K. W.~K.,  {Breivik} K.,  {Kremer} K.,   {Callister} T.,  2020b, arXiv
  e-prints, \href {https://ui.adsabs.harvard.edu/abs/2020arXiv201103564W} {p.
  arXiv:2011.03564}

\bibitem[\protect\citeauthoryear{{Zevin} et~al.,}{{Zevin}
  et~al.}{2020}]{2020arXiv201110057Z}
{Zevin} M.,  et~al., 2020, arXiv e-prints, \href
  {https://ui.adsabs.harvard.edu/abs/2020arXiv201110057Z} {p. arXiv:2011.10057}

\bibitem[\protect\citeauthoryear{{de Souza}, {Yoshida}  \& {Ioka}}{{de Souza}
  et~al.}{2011}]{DeSouza_2011}
{de Souza} R.~S.,  {Yoshida} N.,   {Ioka} K.,  2011, \mn@doi [\aap]
  {10.1051/0004-6361/201117242}, \href
  {http://adsabs.harvard.edu/abs/2011A%26A...533A..32D} {533, A32}

\makeatother
\end{thebibliography}

\end{document}